\documentclass[aps,twocolumn,showpacs,byrevtex,prl,reprint,nofootinbib]{revtex4-2}
\usepackage{epsfig}
\usepackage{dcolumn}
\usepackage{bm}
\usepackage{color}
\usepackage{ulem}
\usepackage{graphics}
\graphicspath{{graphs/}}
\usepackage{subfigure}
\usepackage{overpic}		
\usepackage{times} 
\usepackage[english]{babel}
\usepackage{amsmath}
\usepackage{amssymb}
\usepackage{mathrsfs}	
\usepackage{breqn}		
\usepackage[colorlinks,linkcolor=blue,citecolor=blue]{hyperref}
\usepackage{lineno}

\newcommand{\BESIIIorcid}[1]{\href{https://orcid.org/#1}{\hspace*{0.1em}\raisebox{-0.45ex}{\includegraphics[width=1em]{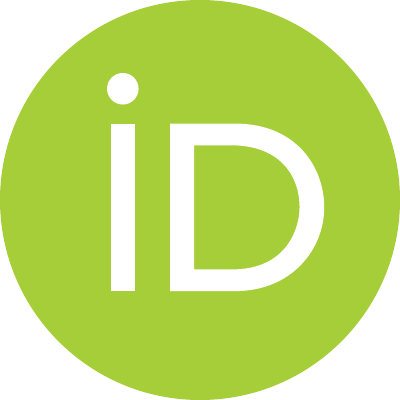}}}}

\addto{\captionsenglish}{%
  
   } \begin{document} \normalsize
  \parskip=5pt plus 1pt minus 1pt
\title{\boldmath Evidence of transverse polarization of $\Xi^0$ hyperon in $\psi(3686)\rightarrow\Xi^0\bar{\Xi}^0$}
\author{BESIII Collaboration}
\thanks{Full author list given at the end of the Letter}

\begin{abstract}
Using $(2.712\pm0.014)\times10^{9}$ $\psi(3686)$ events collected with the BESIII detector at the BEPCII collider, we report an evidence of $\Xi^{0}$ transverse polarization with a significance of 4.4$\sigma$, a precise measurement of the branching fraction and the ratios between the $S$-wave and $D$-wave contributions of $\psi(3686)\to\Xi^{0}\bar\Xi^{0}$.
The weak decay parameters ($\phi_{\Xi^0/\bar{\Xi}^{0}}$, $\alpha_{\Xi^0/\bar{\Xi}^{0}}$) and the angular distribution ($\alpha_{\psi}$) are also measured with higher precision compared to the previous measurements. Furthermore, two the $CP$ observables are also determined to be $A^{\Xi^0}_{CP} = -0.014 \pm 0.030 \pm 0.010$ and
$\Delta\phi^{\Xi^0}_{CP} = 0.000 \pm 0.028 \pm 0.003$~rad, which are still consistent with $CP$ conservation at 1$\sigma$ level under the current statistics.
\end{abstract}
\maketitle
\newpage

The study of hyperon pair production in charmonium decay that occurs in $e^+e^-$ annihilations is an ideal laboratory for exploring hyperon properties such as their polarization and decay, and for conducting tests of fundamental symmetries.
The hyperon-spin polarization in a vector-charmonium decay into a baryon-antibaryon pair was initially explored in Ref.~\cite{Gonzalez:1994zc}. Such an effect requires complex decay amplitudes with a significant relative phase $\Delta\Phi$. To effectively observe the hyperon polarization, the hyperon spin is projected onto a chosen quantization axis, with different spin components manifesting distinct angular dependences in $\cos\theta_{\Xi^{0}}$. Such situation occurs when the state is produced in $e^+e^-$ annihilation with unpolarized beams. It lies in the fact that the photon and the subsequent charmonium can only have $\pm1$ helicities. This specific characteristic provides a potential avenue for the study of the hyperon polarization phenomenon.
The BESIII Collaboration has conducted numerous measurements of the spin and polarization of $\Xi$, $\Lambda$, and $\Sigma$~\cite{BESIII:2016ssr, BESIII:2017kqw, BESIII:2016nix,BESIII:2020fqg, BESIII:2023sgt, BESIII:2024nif, BESIII:2023lkg, BESIII:2018cnd, BESIII:2020fqg, BESIII:2021ypr,BESIII:2021cvv,BESIII:2022qax, BESIII:2022lsz,  BESIII:2023sgt, BESIII:2023drj,BESIII:2023euh,BESIII:2023jhj, BESIII:2024nif, BESIII:2024dmr,BESIII:1530_zhang,BESIII:xi0_zhang,BESIII:lambda_hu},
and finds opposite transverse polarization of the $\Sigma^0$ between $J/\psi$ and $\psi(2S)$ decays~\cite{BESIII:2016ssr, BESIII:2017kqw, BESIII:2016nix,BESIII:2020fqg, BESIII:2023sgt, BESIII:2024nif}, clear transverse polarization of the $\Xi^0$ hyperon in $J/\psi$ decay~\cite{BESIII:2023drj} and no transverse polarization of $\Xi^0$ in $\psi(3686)$ decay~\cite{BESIII:2023lkg}, while no such effect has been observed in other hyperons in $J/\psi$ and $\psi(3686)$ decays~\cite{
BESIII:2018cnd, BESIII:2020fqg, BESIII:2021ypr,BESIII:2021cvv,BESIII:2022qax, BESIII:2022lsz,  BESIII:2023sgt, BESIII:2023drj,BESIII:2023euh,BESIII:2023jhj, BESIII:2024nif, BESIII:2024dmr}.
This has led to a renewed interest in both theoretical and experimental investigations in hyperon physics~\cite{BESIII:2016ssr,BESIII:2017kqw, BESIII:2016nix,BESIII:2020fqg, BESIII:2023sgt,BESIII:2024nif,BESIII:2023lkg, BESIII:2018cnd, BESIII:2020fqg,BESIII:2021ypr,BESIII:2021cvv,BESIII:2022qax, BESIII:2022lsz,BESIII:2023sgt, BESIII:2023drj,BESIII:2023euh,BESIII:2023jhj,BESIII:2024nif, BESIII:2024dmr,Alekseev:2018qjg,Ferroli:2020mra,Wu:2021yfv}. Although these studies provide essential insights into the spin and polarization properties of hyperons, a complete understanding of these phenomena is still lacking.
Taking advantage of the large data samples of $J/\psi$ and $\psi(3686)$ as well as the excellent performance of the BESIII detector, comprehensive measurement of hyperon-antihyperon pair decays can be performed, providing valuable insights into these phenomena.


One of the fundamental open questions in modern physics is the matter-antimatter asymmetry in the Universe. A crucial ingredient for generating this imbalance is the violation of $CP$ symmetry~\cite{CPsource,Liu:2023xhg}. $CP$ violation ($CPV$) has been firmly established in meson decays~\cite{KCP,BCP1,BCP2,DCP}, whose sources are insufficient to account for the observed matter-antimatter asymmetry. In the Standard Model (SM), the Cabibbo-Kobayashi-Maskawa mechanism predicts hyperon $CPV$ effects at the $10^{-4}\sim10^{-5}$ level~\cite{Donoghue:1985ww,xghe,Donoghue:1986hh}, making hyperon decays 
an important system for studying $CPV$. The BESIII Collaboration has recently conducted a series of studies on $CPV$ in hyperon-antihyperon pairs~\cite{BESIII:2018cnd, BESIII:2020fqg, BESIII:2021ypr, BESIII:2022qax, BESIII:2022lsz, BESIII:2023lkg, BESIII:2023sgt, BESIII:2023drj,BESIII:2023jhj, BESIII:2024nif,BESIII:review_zhang,BESIII:1530_zhang,BESIII:xi0_zhang,BESIII:lambda_hu}. Although no evidence for $CPV$ has been observed so far, hyperon decays are still expected to be sensitive to $CPV$ effects~\cite{BESIII:2023sgt,BESIII:2024nif,BESIII:2023drj,BESIII:2022qax}.

In this paper, we present an evidence of the spin polarization of the $\Xi^0$ hyperon and a precise measurement of the branching fraction for $\psi(3686) \rightarrow \Xi^0\bar{\Xi}^0$ through a multidimensional angular distribution analysis. The data statistics, corresponding to $(2.712 \pm 0.014) \times 10^9$ $\psi(3686)$ events~\cite{BESIII:2024lks} collected with the BESIII detector~\cite{Wang:2007tv} at the BEPCII collider~\cite{BESIII}, are approximately five times that used in the previous study~\cite{BESIII:2023lkg}.

\begin{figure}[!htp]
	\centering
	\includegraphics[scale=0.35]{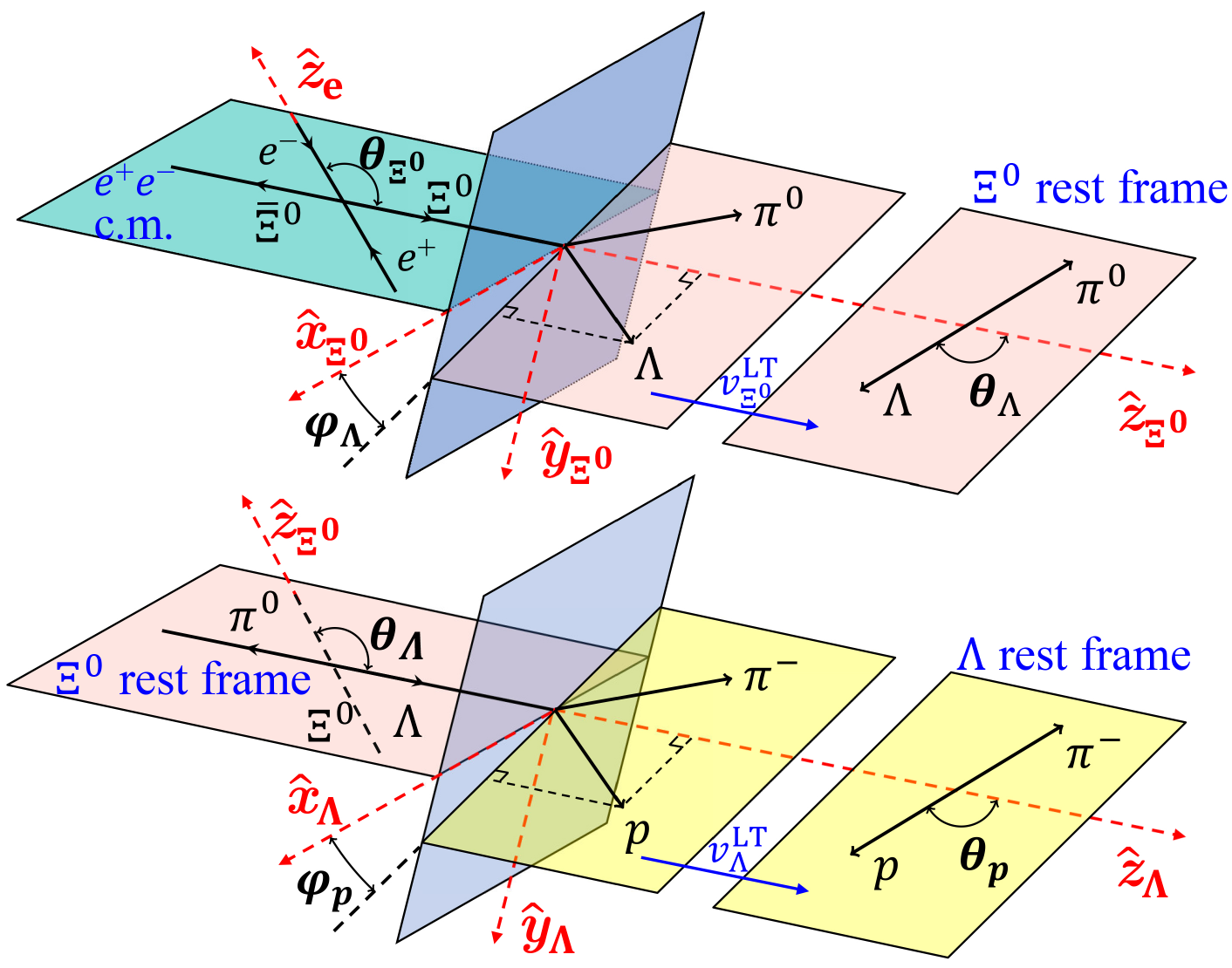}
	\caption{Definition of the coordinate system used to describe the $\psi(3686) \rightarrow \Xi^0\bar{\Xi}^0$ process.
    In the $e^+e^-$ rest frame, the $\hat{\boldsymbol{z}}_{e}$ axis points in the direction of the positron, while the $\hat{\boldsymbol{z}}_{\Xi^0}$ axis is aligned with the momentum direction of the $\Xi^0$.
    In the $\Xi^0$ rest frame, the polar axis is directed along $\hat{\boldsymbol{z}}_{\Xi^0}$, the $\hat{\boldsymbol{y}}_{\Xi^0}$ axis is defined as the cross product $\hat{\boldsymbol{z}}_{e} \times \hat{\boldsymbol{z}}_{\Xi^0}$, and $\hat{\boldsymbol{z}}_{\Lambda}$ is oriented in the direction of the momentum of the $\Lambda$.
    For the $\Lambda$ rest frame, the polar axis is $\hat{\boldsymbol{z}}_{\Lambda}$, with the $\hat{\boldsymbol{y}}_{\Lambda}$ axis given by the cross product $\hat{\boldsymbol{z}}_{\Xi^0} \times \hat{\boldsymbol{z}}_{\Lambda}$.
    The vector $\hat{\boldsymbol{P}}_{\Xi^0} \times \hat{\boldsymbol{z}}_{\Lambda}$ runs parallel to the $\hat{\boldsymbol{y}}_{\Lambda}$ axis. The definition for the $\bar{\Xi}^0$ polarization vector is analogous, with the $\hat{\boldsymbol{z}}_{\bar{\Xi}^0}$ axis against the $\hat{\boldsymbol{z}}_{\Xi^0}$ direction.
    The quantity $v_{\Xi^0/\Lambda}^{\rm LT}$ signifies the Lorentz transformation (LT) corresponding to the velocity $v_{\Xi^0/\Lambda}$.
          }
	\label{xyz}
\end{figure}

To describe the reaction of $e^+ e^- \to \psi(3686) \to \Xi^0 \bar{\Xi}^0 \to \pi^0 \pi^0 \Lambda \bar{\Lambda}$, it is essential to acquire information about each particle in the coordinate system of the parent particle. As shown in Fig.~\ref{xyz}, this employs a right-handed coordinate system, aligning with that provided in Ref.~\cite{formula} to describe the hyperon decay and the orientation of $p/\bar{p}$.
The joint angular distribution function is described by
the vector $\boldsymbol{\xi} =
(\theta_\Xi,\theta_\Lambda,\varphi_\Lambda,\theta_{\bar{\Lambda}},\varphi_{\bar{\Lambda}},\theta_{p},\varphi_{p},\theta_{\bar{p}},\varphi_{\bar{p}})$ as defined in Fig.~\ref{xyz}
and $\boldsymbol{\Omega}$ as
\begin{equation}
	\label{Formula}
	\mathcal{W} = \mathcal{W}(\boldsymbol{\xi},\boldsymbol{\Omega}) = \sum^{3}_{\mu,\bar{\nu}=0} \sum^{3}_{\mu'=0} \sum^{3}_{\bar{\nu}'=0} C_{\mu \bar{\nu}} a^{\Xi}_{\mu \mu'} a^{\Lambda}_{\mu' 0} a^{\bar{\Xi}}_{\bar{\nu} \bar{\nu}'} a^{\bar{\Lambda}}_{\bar{\nu}' 0},
\end{equation}
where $C_{\mu \bar{\nu}}$ denotes the production spin density matrix, $a_{\mu\nu}$ represents the joint decay amplitude, and the set of decay parameters, denoted as
$\boldsymbol{\Omega}$ = $(\alpha_{\psi(3686)}, \Delta \Phi, \alpha_{\Xi^0}, \phi_{\Xi^0}, \alpha_{\Lambda}, \alpha_{\bar{\Lambda}}, \alpha_{\bar{\Xi}^0}, \phi_{\bar{\Xi}^0})$. Detailed definitions of both $C_{\mu \bar{\nu}}$ and $a_{\mu \nu}$ can be found in Ref.~\cite{formula}.

The polarization observable $P_y$ is defined as follows~\cite{Faldt:2017kgy}:
\begin{dmath}
    \label{mu}
     P_y=\frac{\sqrt{1 - \alpha_{\psi(3686)}^2 } \sin 2\theta_\Xi \sin \Delta \Phi}{2(1 + \alpha_{\psi(3686)}\cos^2 \theta_\Xi)},
\end{dmath}
which is dependent on the transverse polarization parameters $\alpha_{\psi}$ and $\Delta \Phi$, and the spin correlations $C_{i j} (i,j=x,y,z\ \text{and}\ C_{0y}\rightarrow P_y)$, are related to the $C_{\mu \bar{\nu}}$ in Eq.~(\ref{Formula}) as
\begin{dmath}
\label{Cmu_nu}
C_{\mu\bar{\nu}} = \left(1 + \alpha_{\psi(3686)} \cos^2 \theta_\Xi\right)
\begin{pmatrix}
1 & 0 & P_y & 0 \\
0 & C_{xx} & 0 & C_{xz} \\
-P_y & 0 & C_{yy} & 0 \\
0 & -C_{xz} & 0 & C_{zz}
\end{pmatrix}.
\end{dmath}
The two $CP$ asymmetry observables $A^{\Xi}_{CP}$, $\Delta \phi^{\Xi}_{CP}$ are defined by
\begin{align}
    \label{eq:CPV}
    &A^{\Xi}_{CP} = \frac{\alpha_{\Xi}+\bar{\alpha}_{\Xi}}{\alpha_{\Xi}-\bar{\alpha}_{\Xi}},\\
    &\Delta\phi^{\Xi}_{CP} = \frac{\phi_{\Xi}+\bar{\phi}_{\Xi}}{2}.
\end{align}

Candidate events are selected by fully reconstructing the subsequent decays of $\Xi^0(\bar{\Xi}^0) \to \pi^0\Lambda(\pi^0\bar{\Lambda})$, $\Lambda(\bar{\Lambda}) \to p \pi^-(\bar{p}\pi^+)$ and $\pi^0 \to \gamma \gamma$. To estimate the detection efficiency, a Monte Carlo (MC) sample of $3 \times 10^7$ simulated $\psi(3686)\to\Xi^0\bar\Xi^0$ events is generated according to phase space (PHSP) using the~\texttt{KKMC} generator~\cite{kkmc1,kkmc2}, where the subsequent decays for $\Xi^0$ and $\bar\Xi^0$ are also generated with the PHSP model. The potential backgrounds are investigated by analyzing the inclusive MC sample of $\psi(3686)$ with the same size as the real data, where the production of the $\psi(3686)$ resonance is simulated with the~\texttt{KKMC} generator, the subsequent decays are processed via~\texttt{EvtGen}~\cite{evtgen1, evtgen2} in accordance with the measured branching fractions provided by the Particle Data Group (PDG)~\cite{PDG2024}, and the remaining unmeasured decay modes are generated with \texttt{Lundcharm}~\cite{lundcharm1,lundcharm2}. The response of the BESIII detector is modeled with MC simulations using a framework based on \texttt{Geant4}~\cite{gent4, gent4soft}.


Charged tracks are required to be reconstructed in the multilayer drift chamber within its angular coverage of $|\cos\theta| < 0.93$, where $\theta$ represents the polar angle with respect to the positron beam direction. Particle identification is carried out based on the track momentum.
The tracks where the momentum is greater than $0.5\,\mathrm{GeV}\!/c$ are considered to be protons and as pions otherwise. Events with at least one $\pi^+$, one $\pi^-$, one $p$ and one $\bar{p}$ are retained for further analysis.

To reconstruct $\Lambda$ and $\bar{\Lambda}$ candidates, a secondary vertex fit is applied to all $p\bar{p}\pi^+\pi^-$ combinations. The best $\Lambda\bar{\Lambda}$ combination is selected by the minimal value of $\sqrt{(M_{p\pi^{-}} - m_{\Lambda})^{2} + (M_{\bar{p}\pi^{+}} - m_{\Lambda})^{2}}$, where $M_{p\pi^{-}}$ ($M_{\bar{p}\pi^{+}}$) is the invariant mass of the $p\pi^{-}$ ($\bar{p}\pi^{+}$) pair and $m_\Lambda$ is the $\Lambda$ mass taken from the PDG~\cite{PDG2024}. The $p\pi^-$ ($\bar{p}\pi^+$) invariant mass is required to be within 5~MeV/$c^2$ of the nominal $\Lambda$ ($\bar{\Lambda}$) mass, which is determined by optimizing the figure of merit (FOM) $S/\sqrt{S + B}$ based on the MC simulation, where $S$ represents the number of signal MC events and $B$ is the number of the background events derived from the inclusive simulation. To further suppress non-$\Lambda$ background, the decay length of $\Lambda(\Bar{\Lambda})$ is required to be greater than zero, where the decay length refers to the distance between its production and decay positions.

Photons are reconstructed from isolated showers in the electromagnetic calorimeter (EMC). The energy deposited in the nearby time of flight counter is incorporated to enhance the reconstruction efficiency and energy resolution. The energies of photons are required to be greater than 25~MeV in the EMC barrel region ($|\cos{\theta}| < 0.8$), and greater than 50~MeV in the EMC end-cap region ($0.86 < |\cos{\theta}| < 0.92$). Moreover, to suppress electronic noise and showers not related to the event, the difference between the EMC time and the event start time is required to fall within the range of ($0, 700$)~ns. Events with at least four photons are retained for further analysis.

In order to further suppress the potential backgrounds and enhance the mass resolution, a six-constraint (6C) kinematic fit is implemented for all $\gamma\gamma\gamma\gamma\Lambda\bar\Lambda$ combinations, imposing energy-momentum conservation from the initial $e^+e^-$ to the selected final states, combined with two additional constraints to the mass of two pairs of photons to form the two $\pi^0$ candidates. Events with $\chi^{2}_{6{\rm C}} < 200$ are retained based on the FOM optimization.

\begin{figure}[!htp]
\centering \includegraphics[scale=0.4]{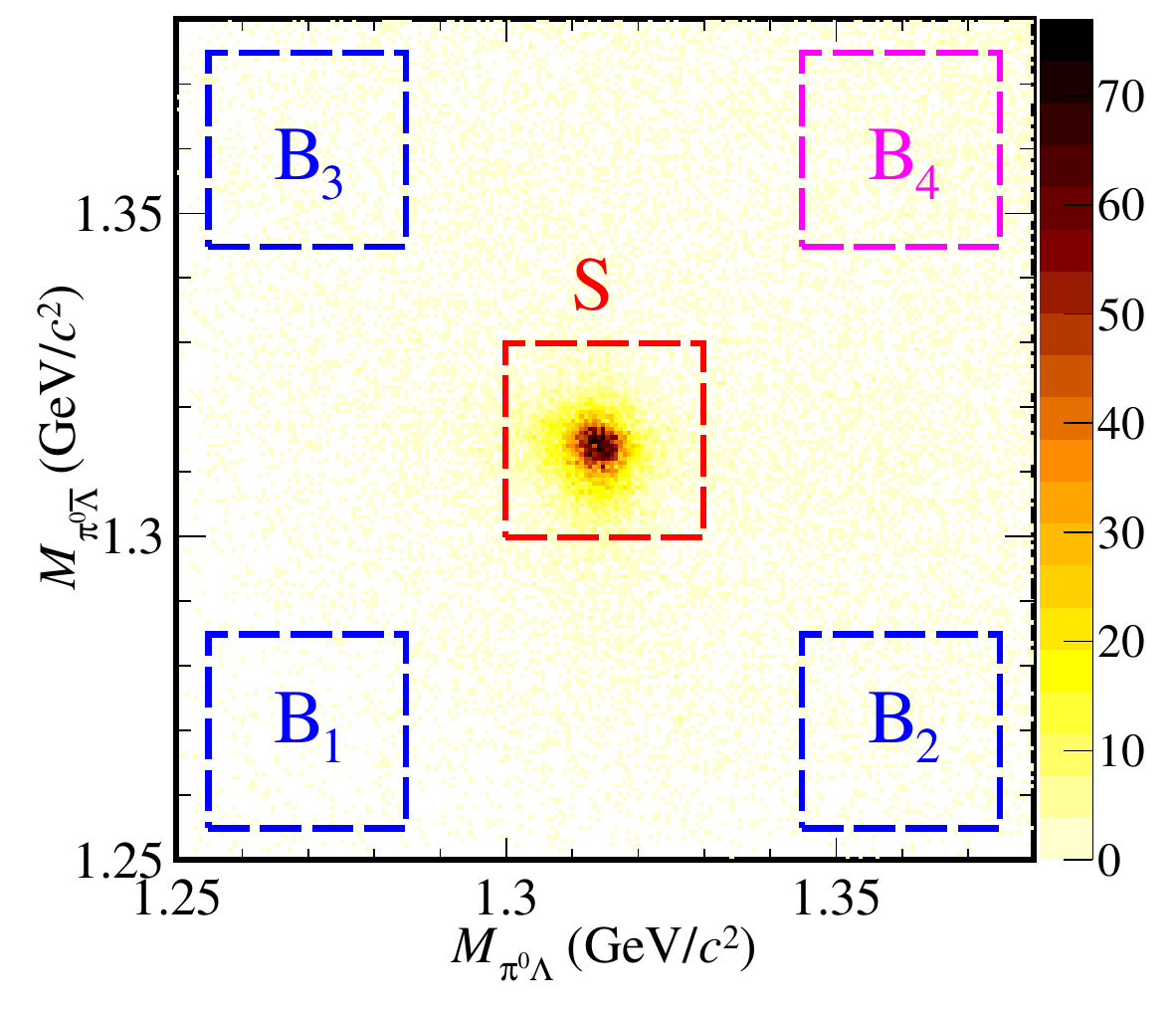}
\caption{Distribution of $ M_{\pi^0 \Lambda} $ versus $ M_{\pi^0 \bar{\Lambda}} $, where the red box ($S$) indicates the signal region and the blue boxes $ \text{B}_i (i = 1, 2, 3) $ represent the selected sideband regions, while the magenta box $\text{B}_4$ is close to the region of the $\Sigma^{*0}\bar\Sigma^{*0}$ background and is not utilized. }
\label{box}
\end{figure}

To reconstruct $\Xi^0$ and $\bar{\Xi}^0$ candidates, the one characterized by the mass minimization, defined as the quantity $\sqrt{(M_{\pi^0\Lambda} - m_{\Xi^0})^2 + (M_{\pi^0 \bar{\Lambda}} - m_{\Xi^0})^2}$, is utilized for all $\pi^0 \pi^0 \Lambda \bar{\Lambda}$ combinations. Here, $M_{\pi^0 \Lambda}$ ($M_{\pi^0 \bar{\Lambda}}$ ) represents the invariant mass of the $\pi^0 \Lambda$ ($\pi^0 \bar{\Lambda}$), and $m_{\Xi^0}$ ($m_{\bar{\Xi}^0}$) is the $\Xi^0$ mass taken from the PDG~\cite{PDG2024}. Figure~\ref{box} presents the distribution of $M_{\pi^0\Lambda}$ versus $M_{\pi^0\bar{\Lambda}}$ based on the above event selection. A distinct enhancement around the known $\Xi^0$ ($\bar{\Xi}^0$) mass can be observed. The $\pi^0 \Lambda$ ($\pi^0 \bar{\Lambda}$) invariant mass is required to be within~15~MeV/$c^2$ of the known $\Xi^0$ ($\bar{\Xi}^0$) mass, which is determined through the FOM.

After applying the above event selection criteria to the data, combined with an event type analysis of the inclusive $\psi(3686)$ MC sample~\cite{TopoAna}, the remaining backgrounds are found to mainly come from $\psi(3686)\rightarrow\pi^0\pi^0J/\psi$ ($J/\psi\to\Lambda\bar\Lambda$), which distributes smoothly in the signal region. The background yield is evaluated by averaging the three sideband regions $\text{B}_i (i=1,2,3)$ as $\sum^{3}_{i=1}\text{B}_{i}/3$ in the $M_{\pi^0\Lambda}$ versus $M_{\pi^0\bar{\Lambda}}$ windows, as depicted in Fig.~\ref{box}:
\begin{align*}
B_1 : [1.2549, 1.2849]\,\mathrm{and}\,[1.2549, 1.2849]~\mathrm{GeV}/c^2,\\
B_2 : [1.3449, 1.3749]\,\mathrm{and}\,[1.2549, 1.2849]~\mathrm{GeV}/c^2,\\
B_3 : [1.2549, 1.2849]\,\mathrm{and}\,[1.3449, 1.3749]~\mathrm{GeV}/c^2.
\end{align*}
The $\text{B}_4$ region is in close proximity to the region of $\Sigma^0(1385) \bar{\Sigma}^0(1385)$ background thus not suitable for background evaluation.
The number of background events $\frac{1}{3}N_{\text{B}_{i}}$ estimated from the aforementioned method is $554 \pm 24$. The number of observed events in data, $N_{obs}=N_{S}-\frac{1}{3}N_{\text{B}_{i}}$, is determined to be 15172 $\pm$ 128. The branching fraction is evaluated to be ${\cal{B}} = \frac{N_{\rm{obs}}}{N_{\psi}\cdot\varepsilon\cdot\mathcal{B}^2(\Xi^0\rightarrow\pi^0\Lambda)\cdot\mathcal{B}^2(\Lambda\rightarrow p\pi^{-})\cdot\mathcal{B}^2(\pi^0\rightarrow\gamma\gamma)} =(2.59 \pm 0.02)\times 10^{-4}$, only with statistical uncertainty. Here $N_{\rm{obs}}$ is the number of observed signal events, $N_{\psi}$ is the number of total $\psi(3686)$ events, and $\varepsilon$ is the detection efficiency incorporating the measured parameters related to the $\Xi^0$ polarization, angular distribution and the $\Xi$ decay parameters listed in Table~\ref{numerical:results:MLL}. $\mathcal{B}$ represents the branching fractions for the subsequent decays $\Xi^0\rightarrow\pi^0\Lambda$, $\Lambda\rightarrow p\pi^{-}$, and $\pi^0\rightarrow\gamma\gamma$, respectively, as provided by the PDG~\cite{PDG2024}.

\begin{figure}[!htbp]
    \centering
    \includegraphics[width=.23\textwidth]{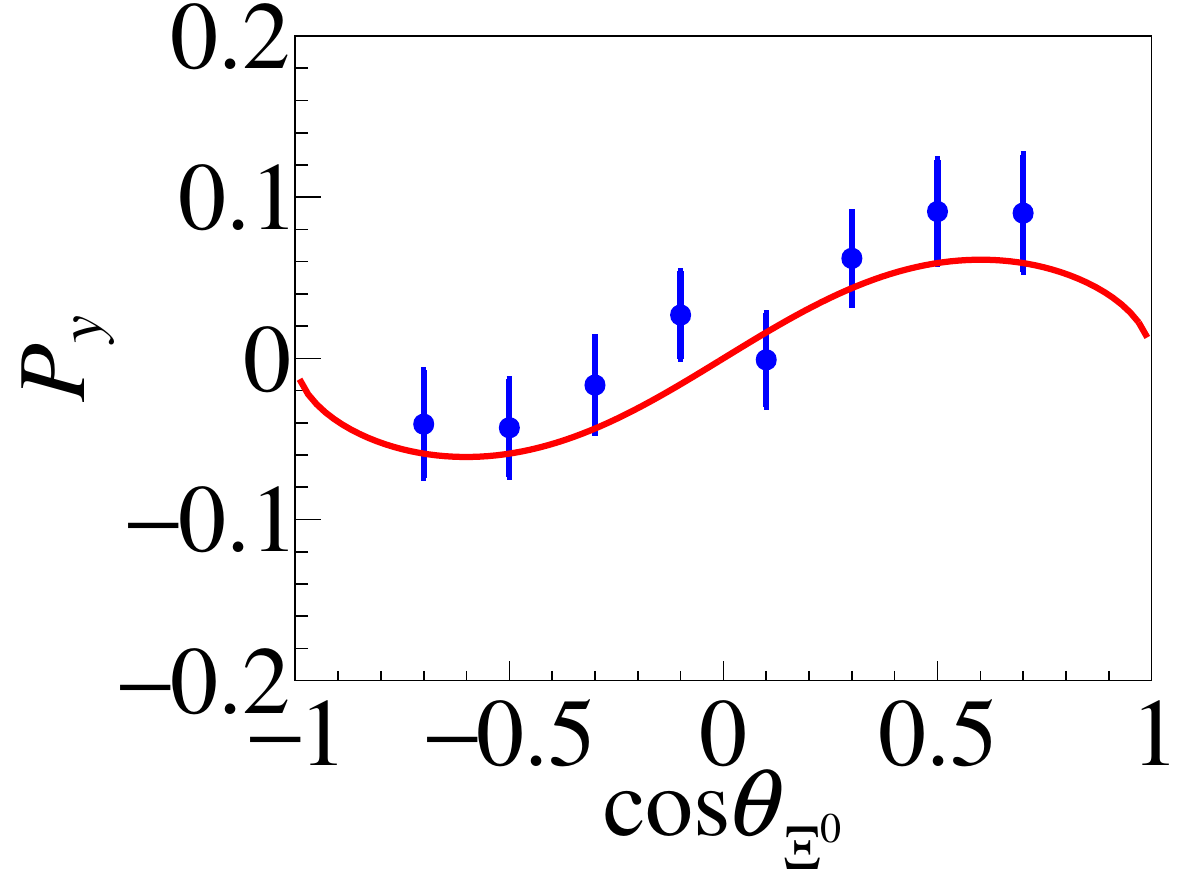}
    \includegraphics[width=.23\textwidth]{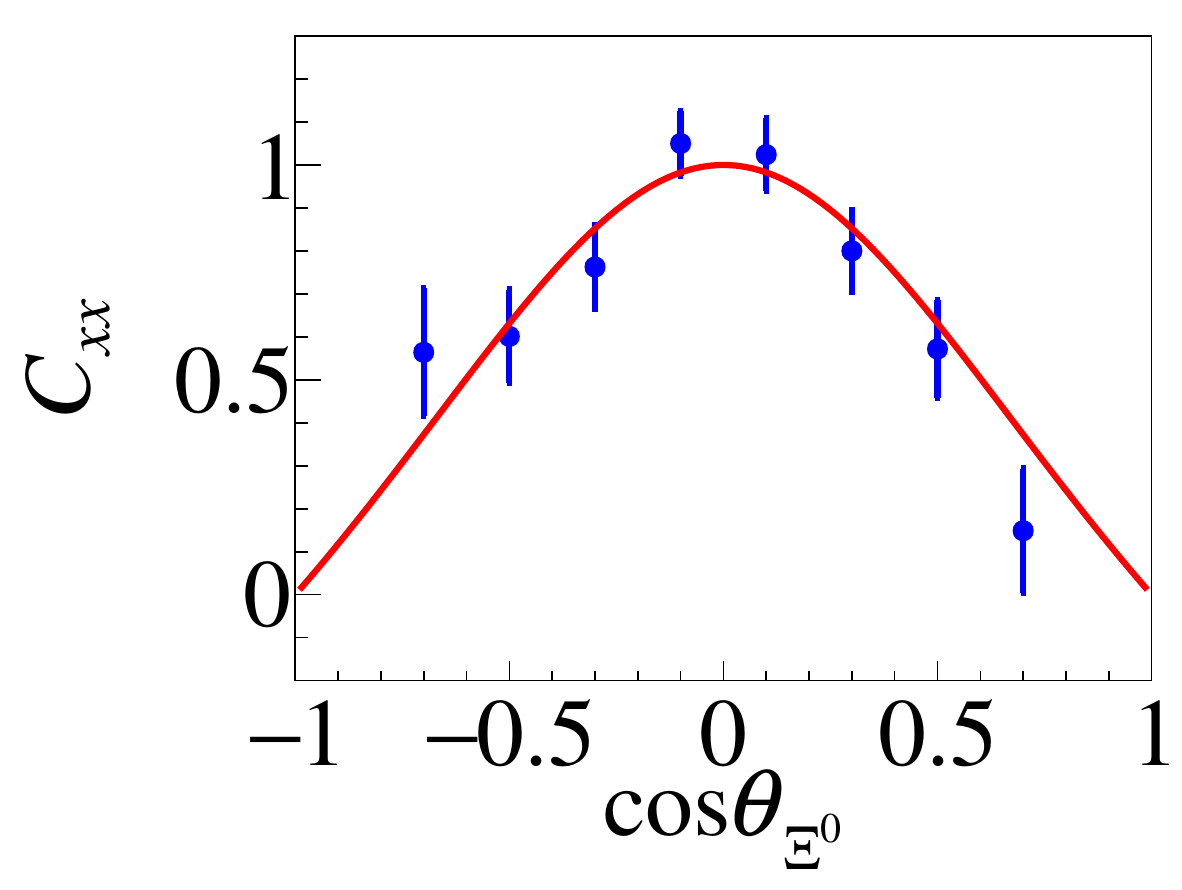}
    \\
    \includegraphics[width=.23\textwidth]{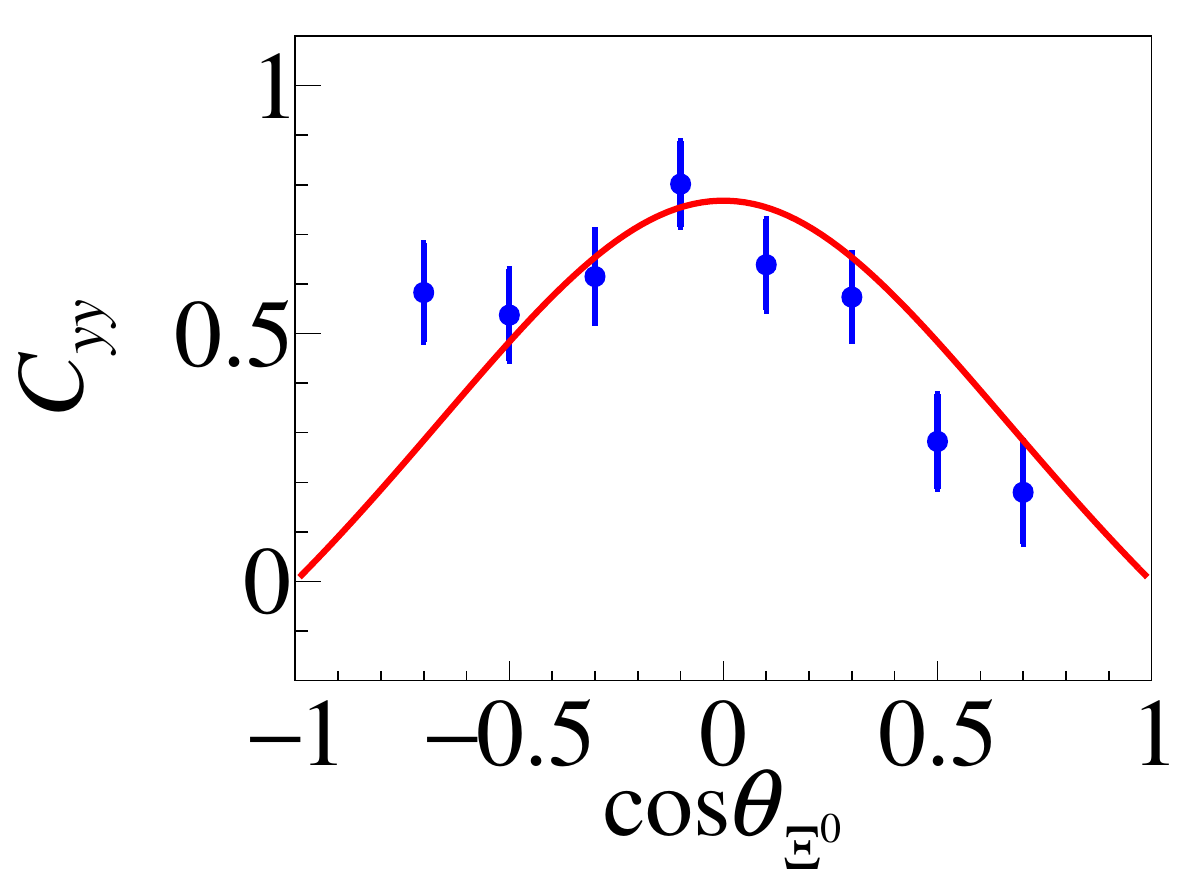}
    \includegraphics[width=.23\textwidth]{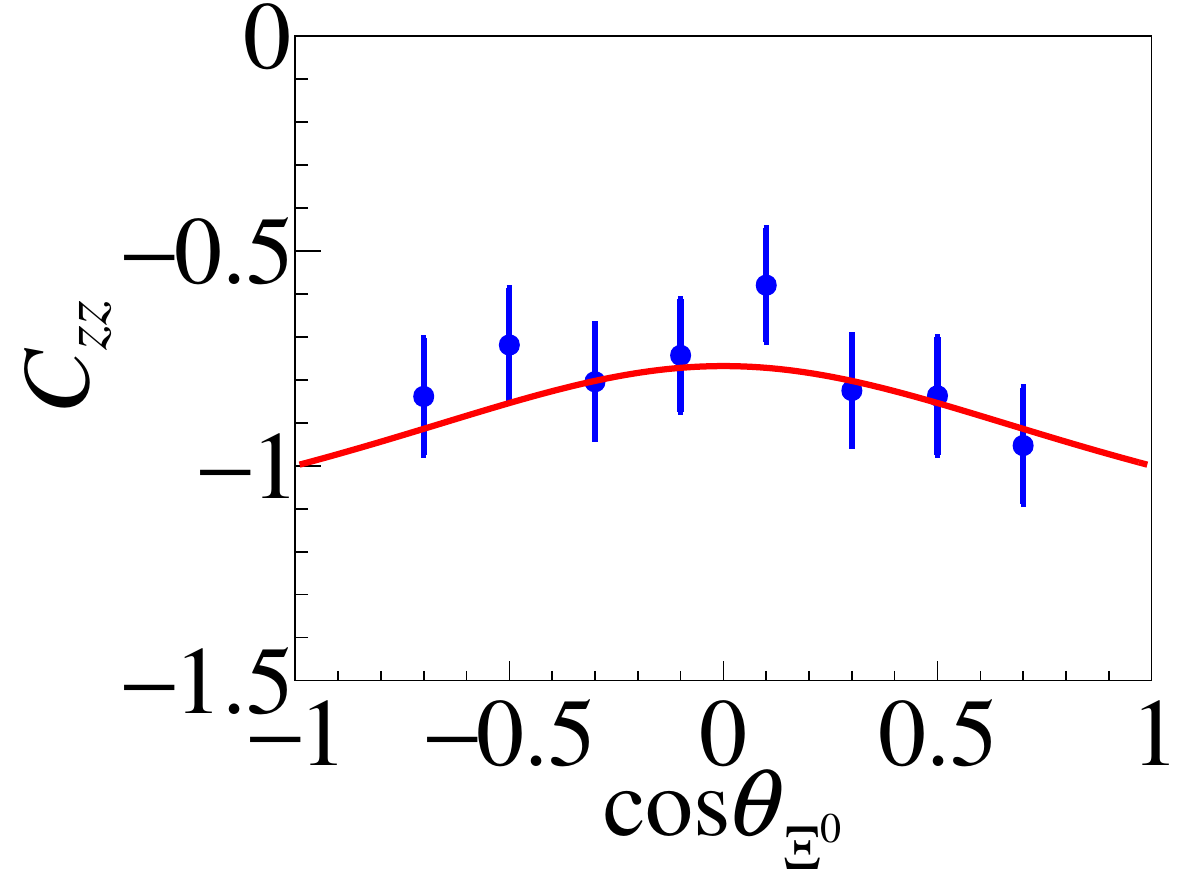}
    \caption{Polarization $P_{y}$ and spin correlations $C_{xx}$,$ C_{yy}$ and $C_{zz}$ in $\psi(3686)\rightarrow \Xi^0\bar{\Xi}^0$ as a function of $\cos\theta_{\Xi^0}$.  The dots with error bars represent the experimental data, and the red line indicates the global fit result. }
    \label{scatter_plot::xxb:projections}
\end{figure}

To determine the set of $\Xi^0$ spin polarization parameters
$\boldsymbol{\Omega}$, an unbinned maximum likelihood fit is implemented.
In the fit, the $\alpha_{\Lambda/\bar\Lambda}$ is fixed at $\pm(\alpha_{\Lambda}-\alpha_{\bar\Lambda})/2=\pm0.754$ by referring to Ref.~\cite{BESIII:2022qax} assuming $CP$ conservation. Here the likelihood function $\mathcal{L}$ is constructed from the probability density function, ${\cal{P}}({\boldsymbol{\xi}}_i)$, for the event $i$ characterized by the measured angles $\boldsymbol{\xi}_i$, as follows:
\begin{align}
    \label{L}
    \mathcal{L} =\prod_{i=1}^{N}\mathcal{P}(\boldsymbol\xi_{i},\mathbf{\Omega})=\prod_{i=1}^{N}\mathcal{C}\mathcal{W}(\boldsymbol\xi_{i},\mathbf{\Omega})\epsilon (\boldsymbol\xi_{i}),
\end{align}
where $N$ is the number of events in the signal region. The joint angular distribution ${\cal{W}}({\boldsymbol{\xi}}_i, {\boldsymbol{\Omega}})$ is given in Eq.~(\ref{Formula}), and $\epsilon(\boldsymbol{\xi}_i)$ is the detection efficiency.
The normalization factor
${\cal{C}}^{-1}=\frac{1}{N_{\rm MC}}\sum_{j=1}^{N_{\rm MC}} {\cal{W}}({\boldsymbol{\xi}}^{j}, {\boldsymbol{\Omega}})$ is calculated as a sum of the corresponding amplitudes $\cal{W}$ from the accepted PHSP MC events $N_{\rm MC}$, applying the same event selection criteria as the data.
The minimization of the objective function defined as
\begin{align}
  \label{S}
\mathcal{S}= -\mathrm{ln}\mathcal{L}_{\rm S} + \frac{1}{3}\mathrm{ln}\mathcal{L}_{\rm B},
\end{align}
is conducted using the \texttt{MINUIT} package from the CERN library~\cite{James:1975dr}.
In Eq.~(\ref{S}), $\mathcal{L}_{\rm S}$ and $\mathcal{L}_{\rm B}$ represent the likelihood function for events chosen in the signal and sideband regions. Figure~\ref{scatter_plot::xxb:projections} shows the resulting polarization parameter $P_{y}$  and spin correlations together with the fit results
with respect to $\cos\theta_{\Xi^0}$ in 10 intervals, where the first and last intervals are ignored due to extremely low statistics.
Table~\ref{numerical:results:MLL} summarizes the numerical results in the fit. The relative phase $\Delta\Phi$ for $\psi(3686)\to\Xi^0\bar\Xi^0$ decay
differs from zero with a significance of larger than 4.4$\sigma$. Other parameters are measured with higher precision and are in agreement with the previous measurement~\cite{BESIII:2023lkg}.

\begin{table}[!htbp] 
    \centering
    \caption{Numerical results for the measured parameters for polarization and angular distribution compared with other measurements~\cite{BESIII:2023lkg,BESIII:2023drj}. Here the first uncertainty is statistical and the second is systematic.}
    \scalebox{0.9}{
    \begin{tabular}{lrrr}
        \hline\hline
        Parameter                   &\multicolumn{1}{c}{This work}    &\multicolumn{1}{c}{Other measurements} &          \\
        \hline
        $\alpha_{\psi(3686)}$	&$0.768\pm0.029\pm0.025$	&$0.665\pm0.086\pm0.081$ &\cite{BESIII:2023lkg}	\\
        $\Delta\Phi$~(rad)	&$0.257\pm0.061\pm0.009$	&$-0.050\pm0.150\pm0.020$&\cite{BESIII:2023lkg}	\\
        $\alpha_{\Xi^0}$	&$-0.345\pm0.015\pm0.003$	&$-0.375\pm0.003\pm0.002$&\cite{BESIII:2023drj}	\\
        $\phi_{\Xi^0}$~(rad)	&$0.008\pm0.041\pm0.002$	&$0.005\pm0.010\pm0.002$&\cite{BESIII:2023drj}	\\
        $\alpha_{\Bar{\Xi}^0}$	&$0.355\pm0.015\pm0.002$	&$0.379\pm0.003\pm0.002$&\cite{BESIII:2023drj}	\\
        $\phi_{\Bar{\Xi}^0}$~(rad)	&$-0.009\pm0.040\pm0.004$	&$-0.005\pm0.010\pm0.002$&\cite{BESIII:2023drj}	\\
        $\xi_{P}-\xi_{S}$~(rad)	&$0.001\pm0.076\pm0.011$	&$0.000\pm0.017\pm0.002$&\cite{BESIII:2023drj}	\\
        $\delta_{P}-\delta_{S}$~(rad)	&$-0.022\pm0.078\pm0.011$	&$-0.013\pm0.017\pm0.004$&\cite{BESIII:2023drj}	\\
        $A^{{\Xi}^0}_{CP}$	&$-0.014\pm0.030\pm0.010$	&$-0.005\pm0.007\pm0.003$&\cite{BESIII:2023drj}	\\
        $\Delta\phi^{\Xi^0}_{CP}$~(rad)	&$0.000\pm0.028\pm0.003$	&$0.000\pm0.007\pm0.001$&\cite{BESIII:2023drj}	\\

        \hline\hline
    \end{tabular}
    }
    \label{numerical:results:MLL}
\end{table}

Systematic uncertainties in the measurement of branching fraction for $\psi(3686)\to\Xi^0\bar\Xi^0$ mainly stem from the differences in detection efficiency between data and MC simulation. The uncertainties related to the $\Xi^0(\bar{\Xi}^0)$ reconstruction efficiency are evaluated using the same strategy as described in Refs.~\cite{BESIII:2019dve, BESIII:2019cuv, BESIII:2021aer, BESIII:2020ktn, BESIII:2021gca, BESIII:2021ccp, BESIII:2022mfx}, and are found to be 5.6\%. The systematic uncertainty due to the 6C kinematic fit is investigated by correcting the helical-track parameters~\cite{Salone:2013helicity} for each charged particle. The difference of 0.2\% from the corrections is regarded as the systematic uncertainty. To evaluate the uncertainty related to the background for sideband, a systematic analysis is carried out by shifting the selected sideband regions ${\rm B}_i (i=1,2,3)$ upwards, to the right, and diagonally by 2~MeV. The largest difference of 0.2\% between new and nominal results is taken as the systematic uncertainty. The uncertainty arising from the continuum background $e^+e^- \to \Xi^0 \bar{\Xi}^0$ is assessed by evaluating the difference with and without considering the continuum process backgrounds, and is found to be 0.3\%.
The uncertainties related to the branching fraction of the intermediate decays $\Xi^0 \rightarrow \pi^0 \Lambda$, $\Lambda \rightarrow p\pi^-$ and $\pi^0 \rightarrow \gamma\gamma$, and the number of $\psi(3686)$ events are quoted to be 0.8\% and 0.4\% from the PDG value~\cite{PDG2024} and Ref.~\cite{BESIII:2024lks}.
The uncertainty related to angular distribution is evaluated
by adjusting the measured parameters of $\Xi^0$ transverse polarization and the spin correlation within $1\sigma$, which induces a maximum difference of 0.2\% and is taken as the systematic uncertainty.
Assuming all sources to be independent, the total systematic uncertainty in the measurement of the branching fraction is determined to be 5.9\% as the sum in quadrature of the mentioned sources.

Systematic uncertainties in the measurement of $\Xi^0$ hyperon polarization and asymmetric decay parameters stem from various sources, such as $\Xi^0$ reconstruction, kinematic fit, background contributions, decay parameters of $\Lambda\to p\pi$, and fit method. The uncertainties related to the $\Xi^0$ reconstruction, including tracking, the requirements on the mass window and decay length of $\Lambda$, are determined using a control sample of $\psi(3686)\to\Xi^0\bar{\Xi}^0$ events. The assessment approach of uncertainties arising from the correction procedure is the same as that in Ref.~\cite{BESIII:2022lsz}, and the discrepancy between the nominal and average values obtained from variations is regarded as the systematic uncertainty.
The uncertainties due to the backgrounds, including the sideband region which represents the non-$\Xi^0(\bar{\Xi}^0)$ process and continuum process, are estimated by comparing the fits with and without considering these background contributions. The systematic uncertainty associated with the 6C kinematic fit is assigned based on the results with and without helical-track parameter corrections~\cite{BESIII:2022lsz}.
The uncertainty resulting from the fixed decay parameters of $\alpha_{\Lambda/\bar\Lambda}$ is estimated by varying the mean values obtained from averaging the results in Ref.~\cite{BESIII:2022qax} within $\pm 1\sigma$. The reliability of the fit method is verified by conducting an input and output check based on 100~pseudoexperiments using the helicity amplitude formula from Ref.~\cite{BESIII:2022qax}. The mean values of polarization and asymmetric decay parameters measured in this analysis are employed as input in the formula, and the number of events in each generated MC sample is 100~times that of the data sample. The differences between the input and output results are assigned as the systematic uncertainties. Assuming all sources to be independent, the total systematic uncertainty is determined by the square root of the quadratic sum. All systematic uncertainties are listed in Table~\ref{SU}.

\begin{table*}[!htbp]
\centering
\caption{\small Systematic uncertainties related to the polarization and asymmetric decay parameters.
Note that the “$\cdots$” indicates that the result is less than $0.001$.
}
\begin{tabular*}{\textwidth}{l@{\extracolsep{\fill}} c c c c c c}
\hline\hline
Program &$\alpha_{\psi(3686)}$  &$\Delta\Phi$   &$\alpha_{\Xi^0}$    &$\phi_{\Xi^0}$    &$\alpha_{\Bar{\Xi}^0}$   &$\phi_{\Bar{\Xi}^0}$\\
\hline The $\Xi^0$ reconstruction                           &$0.023$  &$0.009$  &$0.001$  &$0.001$  &$0.001$  &$0.001$\\
The 6C kinematic fit                                        &$\cdots$  &$0.001$  &$\cdots$  &$0.002$  &$\cdots$  &$0.001$\\
Sideband subtraction                                        &$0.004$  &$0.002$  &$\cdots$  &$\cdots$  &$\cdots$  &$\cdots$\\
Continuum background                                        &$0.006$  &$0.002$  &$\cdots$  &$\cdots$  &$\cdots$  &$\cdots$\\
Decay parameters of $\Lambda$                               &$0.003$  &$0.003$  &$0.001$  &$\cdots$  &$0.001$  &$\cdots$\\
Fit method                                                  &$0.006$  &$\cdots$  &$0.002$  &$0.001$  &$\cdots$  &$0.003$\\
\hline
Total                                                       &$0.025$  &$0.009$  &$0.003$  &$0.002$  &$0.002$  &$0.004$\\
\hline\hline
\end{tabular*}
\label{SU}
\end{table*}

In summary, we report an evidence of spin polarization of the $\Xi^0$ hyperon for the first time and the branching fraction measurement for $\psi(3686) \rightarrow \Xi^0\Bar{\Xi}^0$ using $(2.712 \pm 0.014) \times 10^9$ $\psi(3686)$ events collected with the BESIII detector achieves improved precision over the previous measurements~\cite{BESIII:2023lkg,BESIII:2016nix}.
The relative phase of the psionic form factors~\cite{Faldt:2017kgy} and the branching fraction are measured to be $\Delta\Phi = (0.257 \pm 0.061 \pm 0.010)$ rad, which is found to differ from zero with a significance of $4.4\sigma$, and ${\cal{B}}= (2.59 \pm 0.02 \pm 0.15)\times 10^{-4}$, which is determined with improved precision, and consistent with the previous BESIII measurement~\cite{BESIII:2016nix} within the uncertainty of $1\sigma$. The previous result~\cite{BESIII:2016nix} was based on only a subsample of the data used in the present analysis, and the different tagging and reconstruction strategies introduce systematic uncertainties from different sources. Therefore, the two results can be treated as essentially uncorrelated. Also, it is found that the $\Delta \Phi$ in $\psi(3686)$ decay is obviously different from the one in $J/\psi$ decay ($\Delta \Phi = (1.168 \pm 0.019 \pm 0.018)$ rad)~\cite{BESIII:2023drj}, which could be due to the energy dependence and requires more investigations in theoretical aspects. However, both decays in the $\Xi^0$ sector give the same sign for $\Delta\Phi$, which is different from the one observed in the $\Sigma$ hyperon sector.
Spin-1/2 hyperons produced in a hypreon-antihyperon pair can have either the same or opposite helicity.
In this study, the nonvanishing $\Delta\Phi$, representing the relative phase between the helicity transition amplitudes, is found to be nonzero, indicating a transverse polarization of the $\Xi^0$ hyperon in $\psi(3686)$ decays. 
Besides, the angular distribution parameter for the $\psi(3686) \rightarrow \Xi^0\bar{\Xi}^0$ decay and the weak decay parameters of $\Xi^0$ and $\bar\Xi^0$ are measured with higher precision than those reported in Ref.~\cite{BESIII:2023lkg} and are in good agreement with the previous BESIII measurements~\cite{BESIII:2016nix,BESIII:2023lkg,BESIII:2023drj}.

\begin{table*}[!htbp]
\centering
\caption{Parameters of the $\psi \to \Xi^{-(0)}\bar{\Xi}^{+(0)}$ processes, including the ratio between the coupling constants of $D$ wave and $S$ wave ($g_D/g_S$), the relative phase between $S$ wave and $D$ wave ($\delta$), the ratio between partial width of $S$ wave and total width ($\Gamma_S/\Gamma_{\rm total}$), and the effective radius ($r_{\rm eff}$).
The first and second uncertainties in this paper are statistical and systematic, respectively.}
\label{tab:XiParameters}
\begin{tabular*}{\textwidth}{l@{\extracolsep{\fill}} c c c c}
\hline\hline
Mode & $g_D/g_S~(\mathrm{GeV}^{-2})$ & $\delta~(\mathrm{rad})$ & $\Gamma_S/\Gamma_{\rm total}~(\%)$ & $r_{\rm eff}~(\mathrm{fm})$ \\
\hline
$\psi(3686) \to \Xi^{0}\bar{\Xi}^{0}$ & $0.126 \pm 0.008 \pm 0.005$ & $-0.180 \pm 0.060 \pm 0.021$ & $86.5 \pm 1.4 \pm 1.0$ & $0.041 \pm 0.004 \pm 0.003$ \\
$J/\psi \to \Xi^{0}\bar{\Xi}^{0}$~\cite{BESIII:2023drj} & $0.460 \pm 0.007 \pm 0.007$ & $-0.821 \pm 0.007 \pm 0.017$ & $74.8 \pm 0.6 \pm 0.6$ & $0.122 \pm 0.041 \pm 0.004$ \\
$\psi(3686) \to \Xi^{-}\bar{\Xi}^{+}$~\cite{BESIII:2022lsz} & $0.138 \pm 0.015 \pm 0.011$ & $-0.490 \pm 0.110 \pm 0.080$ & $84.3 \pm 2.9 \pm 2.1$ & $0.048 \pm 0.009 \pm 0.006$ \\
$J/\psi \to \Xi^{-}\bar{\Xi}^{+}$~\cite{BESIII:2021ypr} & $0.491 \pm 0.017 \pm 0.006$ & $-0.743 \pm 0.015 \pm 0.012$ & $73.4 \pm 1.3 \pm 0.5$ & $0.130 \pm 0.006 \pm 0.002$ \\
\hline\hline
\end{tabular*}
\end{table*}

Moreover, following Ref.~\cite{Wu:2021yfv}, with the measured parameters $\alpha_{\psi}$ and $\Delta\Phi$, the ratio of the $D$-wave to $S$-wave coupling constants $g_D/g_S$, the relative phase $\delta$ between the $S$ and $D$ waves, the fraction of the $S$-wave contribution $\Gamma_S/\Gamma_{\mathrm{total}}$, and the effective radius $r_{\mathrm{eff}}$ of the decay $\psi(2S)\to \Xi^{0}\bar{\Xi}^{0}$ are determined for the first time. Here, $\bar{L} = r_{\mathrm{eff}} \times p$, where $\bar{L}$ and $p$ denote the average orbital angular momentum and the relative momentum of the $\Xi^{0}$ and $\bar{\Xi}^{0}$ in the $\psi(2S)$ center-of-mass system, respectively. The results are summarized in Table~\ref{tab:XiParameters}, together with the corresponding parameters for $\psi \to \Xi^{-(0)}\bar{\Xi}^{+(0)}$, which are obtained for the first time following the same procedure~\cite{Wu:2021yfv} and using the values of $\alpha_{\psi}$ and $\Delta\Phi_{\psi}$ from Refs.~\cite{BESIII:2023drj,BESIII:2022lsz,BESIII:2021ypr}. 
Notably, the differences in $g_D/g_S$ and $r_{\mathrm{eff}}$ between the decays $J/\psi \to \Xi^{0}\bar{\Xi}^{0}$ and $\psi(3686) \to \Xi^{0}\bar{\Xi}^{0}$ are approximately the same as those between $J/\psi \to \Xi^{-}\bar{\Xi}^{+}$ and $\psi(3686) \to \Xi^{-}\bar{\Xi}^{+}$, which differs from the pattern observed in the case of $\Sigma$ baryons~\cite{BESIII:2024nif}.
These results indicate that additional, yet unexplored, mechanisms may contribute to hyperon-antihyperon pair production and merit further study.
These results offer important insight into the decay dynamics of charmonium states and the production mechanism of $\Xi^{0}\bar{\Xi}^{0}$ pairs,and serve as useful input for investigations of excited nucleon resonances.

In addition, according to Ref.~\cite{BESIII:2021ypr}, $A^{\Xi}_{CP}$ is proportional to the product of the weak phase difference ($\xi_P - \xi_S$) and the strong phase difference ($\delta_P-\delta_S$) of the final state interaction. Therefore, in the case of a tiny strong phase difference, $A^{\Xi}_{CP}$ would vanish even if the weak phase difference is nonzero. However, $\Delta\phi^{\Xi}_{CP}$ does not have this problem~\cite{Salone:2022lpt} and is more sensitive than $A^{\Xi}_{CP}$ for testing $CPV$, because the strong-phase difference is found to be small for this process (see Table~\ref{numerical:results:MLL}). Two $CP$ observables, $A^{\Xi^0}_{CP}$ and $\Delta\phi^{\Xi^0}_{CP}$, as well as strong and weak phase differences, $\delta_{P}-\delta_{S}$ and $\xi_{P}-\xi_{S}$, are determined and reported in Table~\ref{numerical:results:MLL}.
It is found that the current measurements are consistent with the scenario of $CP$ conservation within the uncertainty of $1\sigma$ under the current statistics.
It is expected that the test of $CPV$ will reach sensitivities comparable to the SM prediction at future super $\tau$-charm factories~\cite{sctf1,sctf2}. 
This measurement serves as an opportunity to study the hyperon structure, providing additional insights into production mechanism of hyperon pairs.

The BESIII Collaboration thanks the staff of BEPCII~\cite{bepcweb} and the IHEP computing center for their strong support. This work is supported in part by 
National Key R\&D Program of China under Contracts No. 2023YFA1606000 and No. 2023YFA1606704; 
the Fundamental Research Funds for the Central Universities under Contracts
No. lzujbky-2025-ytA05,  No. lzujbky-2025-it06, and  No. lzujbky-2024-jdzx06;
National Natural Science Foundation of China under Contracts No. 12075107, No. 12247101, 
No. 11635010, No. 11935015, No. 11935016, No. 11935018, No. 12025502, No. 12035009, No. 12035013, No. 12061131003, No. 12192260, No. 12192261, No. 12192262, No. 12192263, No. 12192264, No. 12192265, No. 12221005, No. 12225509, No. 12235017, and No. 12361141819; 
the Natural Science Foundation of Gansu Province No. 22JR5RA389, No.25JRRA799; the ``111 Center'' under Grant No. B20063;
Guangdong Provincial Key Laboratory of Advanced Particle Detection Technology No. 2024B1212010005;
Guangdong Provincial Key Laboratory of Gamma-Gamma Collider and Its Comprehensive Applications No. 2024KSYS001;
the Chinese Academy of Sciences (CAS) Large-Scale Scientific Facility Program; the Strategic Priority Research Program of Chinese Academy of Sciences under Contract No. XDA0480600; CAS under Contract No. YSBR-101; 100 Talents Program of CAS; The Institute of Nuclear and Particle Physics (INPAC) and Shanghai Key Laboratory for Particle Physics and Cosmology; ERC under Contract No. 758462; German Research Foundation DFG under Contract No. FOR5327; Istituto Nazionale di Fisica Nucleare, Italy; Knut and Alice Wallenberg Foundation under Contracts No. 2021.0174 and No. 2021.0299; Ministry of Development of Turkey under Contract No. DPT2006K-120470; National Research Foundation of Korea under Contract No. NRF-2022R1A2C1092335; National Science and Technology fund of Mongolia; Polish National Science Centre under Contract No. 2024/53/B/ST2/00975; STFC (United Kingdom); Swedish Research Council under Contract No. 2019.04595; and the U. S. Department of Energy under Contract No. DE-FG02-05ER41374.

\begin{widetext}
\begin{center}
\small
M.~Ablikim$^{1}$\BESIIIorcid{0000-0002-3935-619X},
M.-N.~Achasov$^{4,b}$\BESIIIorcid{0000-0002-9400-8622},
P.~Adlarson$^{81}$\BESIIIorcid{0000-0001-6280-3851},
X.-C.~Ai$^{86}$\BESIIIorcid{0000-0003-3856-2415},
R.~Aliberti$^{38}$\BESIIIorcid{0000-0003-3500-4012},
A.~Amoroso$^{80A,80C}$\BESIIIorcid{0000-0002-3095-8610},
Q.~An$^{77,63,\dagger}$,
Y.~Bai$^{61}$\BESIIIorcid{0000-0001-6593-5665},
O.~Bakina$^{39}$\BESIIIorcid{0009-0005-0719-7461},
Y.~Ban$^{49,g}$\BESIIIorcid{0000-0002-1912-0374}
% ...
\\
\vspace{0.2cm}
(BESIII Collaboration)\\
\vspace{0.2cm} {\it
$^{1}$ Institute of High Energy Physics, Beijing 100049, People's Republic of China\\
% ...
$^{4}$ Budker Institute of Nuclear Physics SB RAS (BINP), Novosibirsk 630090, Russia\\
% ...
$^{38}$ Johannes Gutenberg University of Mainz, Johann-Joachim-Becher-Weg 45, D-55099 Mainz, Germany\\
$^{39}$ Joint Institute for Nuclear Research, 141980 Dubna, Moscow region, Russia\\
% ...
$^{49}$ Peking University, Beijing 100871, People's Republic of China\\
% ...
$^{61}$ Southeast University, Nanjing 211100, People's Republic of China\\
% ...
$^{63}$ State Key Laboratory of Particle Detection and Electronics, Beijing 100049, Hefei 230026, People's Republic of China\\
% ...
$^{77}$ University of Science and Technology of China, Hefei 230026, People's Republic of China\\
% ...
$^{80}$ University of Turin and INFN, (A)University of Turin, I-10125, Turin, Italy; (B)University of Eastern Piedmont, I-15121, Alessandria, Italy; (C)INFN, I-10125, Turin, Italy\\
$^{81}$ Uppsala University, Box 516, SE-75120 Uppsala, Sweden\\
% ...
$^{86}$ Zhengzhou University, Zhengzhou 450001, People's Republic of China\\

\vspace{0.2cm}
$^{\dagger}$ Deceased\\
% ...
$^{b}$ Also at the Novosibirsk State University, Novosibirsk, 630090, Russia\\
% ...
$^{g}$ Also at State Key Laboratory of Nuclear Physics and Technology, Peking University, Beijing 100871, People's Republic of China\\
% ...
}
\end{center}
\end{widetext}

\end{document}